\documentclass[aps,prc,twocolumn,superscriptaddress,showpacs,nofootinbib]{revtex4-2}

\usepackage{graphicx}% Include figure files
\usepackage{longtable}
\usepackage{dcolumn}% Align table columns on decimal point
\usepackage{bm}% bold math
\usepackage{ulem}
\usepackage{color}
\usepackage{epsfig}
\usepackage{natbib}

\newcommand{\al}{$\alpha$}
\newcommand{\g}{$\gamma$}
\newcommand{\raa}{($\alpha$,$\alpha$)}
\newcommand{\raX}{($\alpha$,$X$)}

\newcommand{\rag}{($\alpha$,$\gamma$)}
\newcommand{\ran}{($\alpha$,n)}

\newcommand{\rann}{($\alpha$,2n)}

\newcommand{\rap}{($\alpha$,p)}

\newcommand{\rga}{($\gamma$,$\alpha$)}

\newcommand{\stot}{$\sigma_{\rm{reac}}$}

\newcommand{\gpro}{$\gamma$-process}

\newcommand{\sfact}{S-factor}

\newcommand{\tetwonull}{$^{120}$Te}
\newcommand{\teii}{$^{122}$Te}
\newcommand{\teiii}{$^{123}$Te}
\newcommand{\teiv}{$^{124}$Te}
\newcommand{\tev}{$^{125}$Te}

\newcommand{\tethreenull}{$^{130}$Te}
\newcommand{\xetwoiii}{$^{123}$Xe}
\newcommand{\xetwoiv}{$^{124}$Xe}
\newcommand{\xetwov}{$^{125}$Xe}

\newcommand{\xetwovii}{$^{127}$Xe}
\newcommand{\xetwoviii}{$^{128}$Xe}
\newcommand{\xethreeiii}{$^{133}$Xe}
\newcommand{\ithreeiii}{$^{133}$I}
\begin{document}

\title{
Low energy $\alpha$-nucleus optical potential studied via ($\alpha$,n) cross section measurements on Te isotopes
}
\author{Zs. M\'atyus}%
\affiliation{Institute for Nuclear Research (ATOMKI), 4001 Debrecen, Hungary}
\affiliation{University of Debrecen, Doctoral School of Physics, Egyetem t\'er 1., 4032 Debrecen, Hungary}
\author{Gy. Gy\"urky}%
\email{corresponding author, gyurky@atomki.hu}
\affiliation{Institute for Nuclear Research (ATOMKI), 4001 Debrecen, Hungary}
\author{P. Mohr}%
\affiliation{Institute for Nuclear Research (ATOMKI), 4001 Debrecen, Hungary}
\author{A. Angyal}%
\affiliation{Institute for Nuclear Research (ATOMKI), 4001 Debrecen, Hungary}
\author{Z. Hal\'asz}%
\affiliation{Institute for Nuclear Research (ATOMKI), 4001 Debrecen, Hungary}
\author{G.G. Kiss}%
\affiliation{Institute for Nuclear Research (ATOMKI), 4001 Debrecen, Hungary}
\author{\'A T\'oth}%
\affiliation{Institute for Nuclear Research (ATOMKI), 4001 Debrecen, Hungary}
\affiliation{University of Debrecen, Doctoral School of Physics, Egyetem t\'er 1., 4032 Debrecen, Hungary}
\author{T. Sz\"ucs}
\affiliation{Institute for Nuclear Research (ATOMKI), 4001 Debrecen, Hungary}
\author{Zs. F\"ul\"op}%
\affiliation{Institute for Nuclear Research (ATOMKI), 4001 Debrecen, Hungary}

\date{\today}

\begin{abstract}
\begin{description}

\item[Background]
In several processes of stellar nucleosynthesis, like the astrophysical $\gamma$-process, nuclear reactions involving $\alpha$ particles play an important role. The description of these reactions necessitates the knowledge of the $\alpha$-nucleus optical model potential (AOMP) which is highly ambiguous at low, astrophysical energies. This ambiguity introduces a substantial uncertainty in the stellar models for predicting elemental and isotopic abundances. 

\item[Purpose]
The experimental study of the AOMP is thus necessary which can be implemented by measuring the cross section of $\alpha$-induced nuclear reactions. At low energies, \ran\ reactions are suitable for such a purpose. Therefore, in the present work, the \ran\ cross sections of four Te isotopes have been measured, mostly for the first time, and compared with theoretical predictions.

\item[Method]
The \ran\ cross sections of $^{120,122,124,130}$Te have been measured in the energy range between about 10 and 17\,MeV using the activation method. The detection of the $\gamma$ radiation following the decay of the radioactive reaction products were used to determine the cross sections.

\item[Results]
The measured cross sections are compared with statistical model calculations obtained from the widely used TALYS nuclear reaction simulation code. Predictions using various available AOMPs are investigated. 

\item[Conclusions]
It is found that the recently developed Atomki-V2 AOMP provides the best description for all studied reactions and this potential also reproduces well the total reaction cross sections from elastic scattering experiments, when they are available in literature. We recommend therefore to use the astrophysical reaction rates based on this potential for nucleosynthesis models of heavy elements.

\end{description}
\end{abstract}

%\pacs{ }

\maketitle

\section{Introduction}
\label{sec:intro}

Stars generate energy and synthesize chemical elements through nuclear
reactions. These reactions also strongly influence the evolution and final
fate of stars. The most important nuclear physics quantity is the reaction
cross section which determines the astrophysical reaction rates in the stellar
interior. On the one hand, reaction rates for nuclear reactions with positive
Q-value are the basis for the energy generation in stars. On the other hand,
neuclear reaction rates define the amount and abundance distribution of
heavier elements produced by the stars in the various stages of their
evolution.

Ideally, the cross sections of astrophysically important reactions should be known experimentally in the energy range relevant for the given stellar process (in the so-called Gamow window \cite{Iliadis2015}). The typical temperatures of stars, however, result in such low energies that direct measurement of the corresponding extremely low cross sections is not possible in most of the cases. Therefore, low energy extrapolations based on theoretical considerations are almost always necessary. Various basic nuclear physics parameters are needed for such a theoretical cross section calculation.

In the wide region of chemical elements above iron, the level density of isotopes is typically so high that statistical treatment in the cross section calculations becomes possible and also often inevitable. This can be carried out by the Hauser-Feshbach approach \cite{Hauser1952}. This method requires several input parameters such as nuclear masses, level densities, $\gamma$-ray strength functions and models describing the interaction between nuclei \cite{Rauscher2011}. 

In various astrophysical processes, reactions involving $\alpha$ particles play an important role. A prime example is the astrophysical $\gamma$-process \cite{Arnould2003,Rauscher2013} which is thought to be the main source of those heavy, proton-rich isotopes (the p-nuclei) which cannot be produced by the neutron capture based s- \cite{Kappeler2011,Lugaro2023} and r-processes \cite{Cowan2021}. The $\gamma$-process proceeds mainly through $\gamma$-induced reactions, and on the proton-rich side of the valley of nuclear stability \rga\ reactions become important. For the theoretical description of the \rga\ reactions in the framework of the Hauser-Feshbach statistical model, the interaction of the residual nucleus and the $\alpha$ particle in the exit channel is based on the $\alpha$-nucleus optical model potential (AOMP). This potential must be known with high accuracy in order to describe precisely the \rga\ reactions and hence to provide reliable cross section data for $\gamma$-process nucleosynthesis models.

Historically, the AOMP was studied mainly via elastic scattering reactions, and this type of reaction is still a powerful tool for such studies \cite{Kiss2022}. In a scattering experiment, however, the cross section must deviate substantially from the Rutherford scattering cross section in order to assess the AOMP. This requires relatively high energies, typically above the astrophysically relevant energy range. One of the widely used AOMPs (the so-called McFadded-Satchler potential, MCF in the following) was developed from E$_\alpha$\,=\,24.7\,MeV scattering data \cite{MCF1966}. When a new motivation from nuclear astrophysics arose and this potential was applied at much lower energies (around 10\,MeV and below), it turned out that it is not able to give a good description of the experimental data. The need for a detailed study of low energy AOMPs was thus formulated \cite{Somorjai1998}.

Several new AOMP models have been created since the MCF potential (see the discussion in Sec.\,\ref{sec:theo}) which often result in largely different cross section predictions for \rga\ reactions. Therefore, further experimental study of the low energy AOMP is necessary in order to find the best potential models and fine-tune their parameters. 

Besides elastic scattering experiments, the AOMP can also be studied by measuring the cross sections of $\alpha$-induced nuclear reactions \cite{Basak2022}. The cross section measurement of radiative capture \rag\ reactions is preferred over the direct study of \rga\ reactions as the effect of thermal excitation is much less severe \cite{Mohr2007,Rauscher2014}. Unfortunately, \rag\ reactions usually have low cross sections in the mass and energy range of the $\gamma$-process which makes the experiments challenging. An alternative approach is the measurement of \ran\ reactions which have higher cross sections above their threshold. Measured \ran\ cross sections can provide information about the AOMP as it was demonstrated recently e.g. in \cite{Gyurky2023}.

In addition to the $\gamma$-process discussed above, \ran\ reactions have direct relevance in the modeling of the weak r-process of nucleosynthesis \cite{Mohr2016,Arcones2011}. Experimental reaction rate data are also needed for a better understanding of this process \cite{Szegedi2021,Psaltis_APJ2022,Kiss_APJ2021_zr96an}. 

The aim of the present work is thus to measure the \ran\ cross sections of four tellurium isotopes for which no experimental data exist in the studied energy range. This paper is organized as follows: after providing some further information about the investigated reactions in Sec.\,\ref{sec:reactions}, details of the experiments are given in Sec.\,\ref{sec:experiment} and the results are presented in Sec.\,\ref{sec:results}. Sec.\,\ref{sec:theo} contains the analysis of the data from the optical model point of view, while Sec.\,\ref{sec:conc} gives a short summary and conclusions.

\section{Investigated reactions}
\label{sec:reactions}

Tellurium has eight stable isotopes with mass numbers and natural abundances of 120 (0.0921\%\,$\pm$\,0.0003\%), 122 (2.529\%\,$\pm$\,0.006\%), 123 (0.884\%\,$\pm$\,0.002\%), 124 (4.715\%\,$\pm$\,0.012\%), 125 (7.048\%\,$\pm$\,0.018\%), 126 (18.798\%\,$\pm$\,0.047\%), 128 (31.74\%\,$\pm$\,0.08\%) and 130 (34.16\%\,$\pm$\,0.09\%). These abundance values are taken from the latest compilation \cite{IUPAC2016} and updated by the very recent precise measurement of Te isotopic ratios \cite{Murugan2020}. \ran\ reactions on these nuclei lead to various Xe isotopes, four of them ($^{123,125,127,133}$Xe) are radioactive. 

Cross section measurement of \ran\ reactions in the mass region of the \gpro\ using neutron detection is extremely difficult due mainly to background problems caused by high cross section, neutron-emitting nuclear reactions on target impurities. Therefore, the activation method \cite{Gyurky2019} is usually preferred which is based on the detection of the decay-radiation of the reaction products. This technique was used in the present work and owing to the four radioactive Xe reaction products, four reaction cross sections were studied: $^{120}$Te\ran$^{123}$Xe, $^{122}$Te\ran$^{125}$Xe, $^{124}$Te\ran$^{127}$Xe and $^{130}$Te\ran$^{133}$Xe.

\begin{table}
\caption{\label{tab:decay} Decay parameters of the Te\ran\ reaction products. The values are taken from the most recent compilations \cite{NDS_A123,NDS_A125,NDS_A127,NDS_A133} with the exception of the $^{125}$Xe half-life, where the more precise results of a recent measurement is quoted \cite{Szegedi2019}.}
\begin{ruledtabular}
\begin{tabular}{llll}
%\hline
Reaction  & Half-life & E$_\gamma$ & Relative \\
          & 					 &	[keV]			&	intensity [\%] \\
\hline
$^{120}$Te\ran$^{123}$Xe & (2.050\,$\pm$\,0.014)\,h &	148.9	&  49.1\,$\pm$\,0.6	\\
                         &                          &	178.1	&  15.5\,$\pm$\,0.7	\\
                         &                          &	330.2	&  8.6\,$\pm$\,0.5	\\
$^{123}$Xe($\beta^+$)$^{123}$I & (13.223\,$\pm$\,0.002)\,h &	159.0	&  83.60\,$\pm$\,0.19	\\										
$^{122}$Te\ran$^{125}$Xe & (16.87\,$\pm$\,0.08)\,h  &	188.4	&  53.8$^*$	\\
                         &                          &	243.4	&  30.0\,$\pm$\,0.6	\\
                         &                          &	453.8	&  4.67\,$\pm$\,0.10	\\
$^{124}$Te\ran$^{127}$Xe & (36.346\,$\pm$\,0.003)\,d &	145.3	&  4.26\,$\pm$\,0.15	\\
                         &                          &	172.1	&  25.4\,$\pm$\,0.9	\\
                         &                          &	202.9	&  67.8\,$\pm$\,1.2	\\
                         &                          &	375.0	&  17.1\,$\pm$\,0.6	\\
$^{130}$Te\ran$^{133}$Xe$^m$ & (2.198\,$\pm$\,0.013)\,d &	233.2	&  10.12\,$\pm$\,0.15	\\
$^{130}$Te\ran$^{133}$Xe$^g$ & (5.2475\,$\pm$\,0.0005)\,d &	79.6	& 0.44\,$\pm$\,0.18	\\
                         &                        &	81.0	&  36.9\,$\pm$\,0.3	\\
\end{tabular}
\end{ruledtabular}
\vspace{-4mm}
\flushleft \footnotesize{$^*$No uncertainty is given in the compilation \cite{NDS_A125}.}
\end{table}

The decay parameters of the reaction products taken from the latest compilations \cite{NDS_A123,NDS_A125,NDS_A127,NDS_A133} are summarized in Table\,\ref{tab:decay}. The $\beta$-decay of the produced Xe isotopes is followed by the emission of characteristic $\gamma$ radiation. The detection of these $\gamma$ rays were used for the cross section determination. The table contains only the most intense $\gamma$ transitions which were used for the analysis. Besides its ground state, $^{133}$Xe has a long-lived isomeric state which has decay signature different from the ground state. Therefore, in the case of the $^{130}$Te\ran$^{133}$Xe reaction, the cross sections leading both to the ground and isomeric states of $^{133}$Xe could be measured.

\section{Experimental procedure}
\label{sec:experiment}

\subsection{Target preparation and characterization}
\label{subsec:target}

In order to measure the \ran\ cross sections of all studied isotopes at a given energy in a single irradiation, natural isotopic composition targets were used. The targets were prepared by vacuum evaporation. High chemical purity ($>$99.999\,\%) metallic Te was evaporated onto 10\,$\mu$m thick Al foil backings. The foils were fixed in annular target holders with 16\,mm in inner diameter. The weight measurement of the Al foils with 1\,$\mu$g precision before and after the evaporation provided the first information about the target thicknesses, which were typically around 600\,$\mu$g/cm$^2$. Besides several test evaporations, eight targets were prepared and characterized as described below. 

Thin layers of Te may oxidize in the residual gas of the vacuum evaporator or after the target preparation. The weight measurement of the targets, therefore, does not provide the Te target thickness (i.e. the surface number density of the Te atoms) as the Te:O stoichiometric ratio is not known a priori. The Te target thicknesses were thus measured with two independent ion beam analysis methods: Rutherford Backscattering Spectrometry (RBS) and Particle-induced X-ray Emission (PIXE).

\begin{figure*}
\centering
\includegraphics[width=\textwidth]{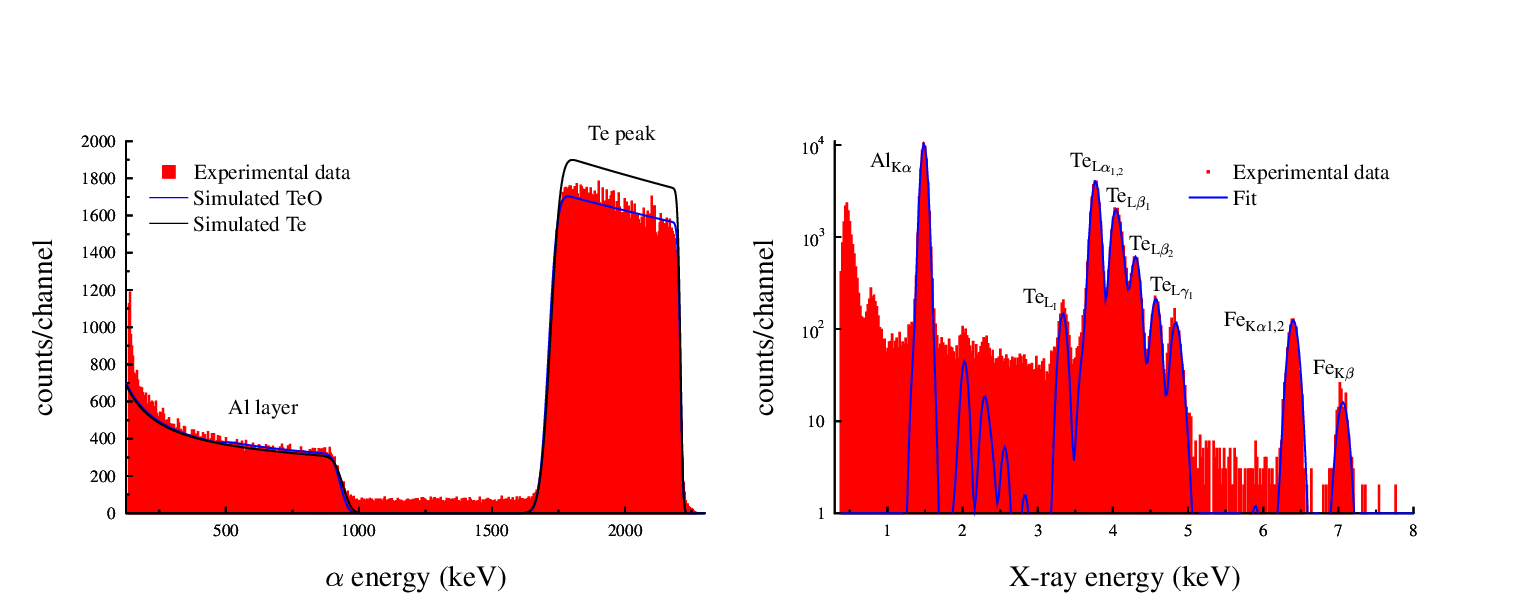}
\caption{
  \label{fig:RBS_PIXE}
  (left) A measured and fitted RBS spectrum. Besides the best fit labeled as TeO, a fit supposing a pure Te layer is also shown (labeled as Te) indicating the need for including oxygen in the layer composition. (right) PIXE spectrum indicating the peaks corresponding to Te, Al and Fe. The latter one is the highest concentration impurity in the Al foil. 
}
\end{figure*}

The RBS measurement was carried out at the Tandetron accelerator of Atomki \cite{Rajta2018}. The targets were bombarded by a 2.5\,MeV $\alpha$ beam with a few hundred nA intensity. The backscattered $\alpha$ particles were detected with an ion implanted Si detector placed at 165 degrees with respect to the beam direction. The acquired spectra were analyzed with the SIMNRA code \cite{SIMNRA} to obtain the Te target thickness and in addition the Te:O ratio. The beam size used for the RBS measurement was similar to the one of the cross section measurement. The left panel of Fig.\,\ref{fig:RBS_PIXE} shows a typical RBS spectrum with the best fit and with a calculated spectrum supposing a pure, oxygen-free Te layer. It can clearly be seen that without supposing oxygen in the layer, the measured spectra cannot be reproduced. The obtained Te:O atomic ratios were typically 0.7:0.3 and Te surface densities were in the range of (2.0 - 2.5)$\cdot$10$^{18}$ atoms/cm$^2$.

The PIXE measurements were also carried out at the Tandetron accelerator using its microbeam setup. The targets were bombarded by 3.2\,MeV protons and the induced X-rays were detected with an SDD detector with AP3.3 ultra thin polymer window (SGX Sesortech). A permanent magnet protected the detector from the scattered protons. The beam dose was measured with a beam chopper \cite{Bartha2000}. PIXE spectra were recorded with an SGX DX200 digital DPP and evaluated with the GUPIXWIN program code \cite{GUPIX} in order to obtain the Te target thickness. Exploiting the capabilities of the microbeam setup, spectra were taken on three different positions on each target with a few mm apart, scanning the beam on a 1\,mm\,$\times$\,1\,mm area. Information on the target homogeneity could be obtained this way. Thicknesses measured at different positions on a given target were always in agreement within the uncertainty. Figure\,\ref{fig:RBS_PIXE} shows a typical
PIXE spectrum where the relevant peaks are indicated (right panel).

The Te thicknesses from the RBS and PIXE measurements were always in agreement, the larges deviation was 4\,\%, well within the 5\,\% uncertainty characteristic for both methods. The average of the two results was thus adopted as the final target thickness used for the cross section determination. 

\subsection{Irradiations}
\label{sec:irrad}

For the \ran\ cross section measurements the targets were irradiated by $\alpha$ beams provided by the cyclotron accelerator of Atomki \cite{Biri2021}. The target chamber was the same as used in our recent experiments, see Fig.\,3. in ref. \cite{Korkulu2018}. The typical He$^{++}$ beam intensity was about 1\,$\mu$A which was measured by a charge integrator. The fluctuations in the beam intensity were taken into account by recording the collected charge in the chamber as a function of time with 1 min time resolution. 

The length of the irradiations varied between 3 and 41 hours. Longer irradiations were used at the lowest energies where the cross sections are the smallest. The studied $\alpha$-energy range between 10 and 17\,MeV was covered with 1\,MeV steps, eight irradiations were thus carried out. The lowest measurable energy was determined by the dropping cross section, below this energy the cross sections could not be determined with reasonable precision. Energies higher than 17\,MeV were not studied, on the one hand because of the decreasing astrophysical relevance and on the other hand because of the opening of other reaction channels, as will be discussed in Sec.\,\ref{sec:results}. 

Due to technical reasons, the cyclotron accelerator could not provide a 10\,MeV $\alpha$ beam. Therefore, the lowest energy point was measured using an energy degrader foil. An Al foil of 10.14 $\mu$m thickness was placed in front of the Te target. The precise thickness of the degrader foil was determined by $\alpha$-energy loss measurement. The spectrum of $\alpha$ particles from a mixed $^{239}$Pu-$^{241}$Am-$^{244}$Cm calibration source after passing through the foil was measured in an alpha spectrometer. Based on the stopping power of Al obtained from the SRIM code \cite{SRIM}, the foil thickness was determined with 4\,\% accuracy. This results was used to obtain the interaction energy at the Te target.

\subsection{Detection of the decay radiation}

After the irradiations, the targets containing the created reaction products were removed from the irradiation chamber and transported to a $\gamma$ detector setup for off-line decay counting. As it can be seen in Table\,\ref{tab:decay}, the decay of the reaction products is mainly followed by low-energy $\gamma$ radiation. Therefore, a planar detector consisting of a Ge crystal with 60.7\,mm diameter and 26.4\,mm thickness and a thin carbon epoxy window was used which has high efficiency for low energy $\gamma$ rays, good energy resolution and - owing to the small crystal size - relatively low background caused by higher energy $\gamma$ radiation. For the first two measurements (at 12 and 17 MeV $\alpha$ energies), however, the planar detector was not available. Therefore, for these runs a 100\,\% relative efficiency coaxial HPGe detector was used. Both detectors were put in complete 4\,$\pi$ lead shielding against laboratory background.

\begin{figure}
\centering
\includegraphics[width=\columnwidth]{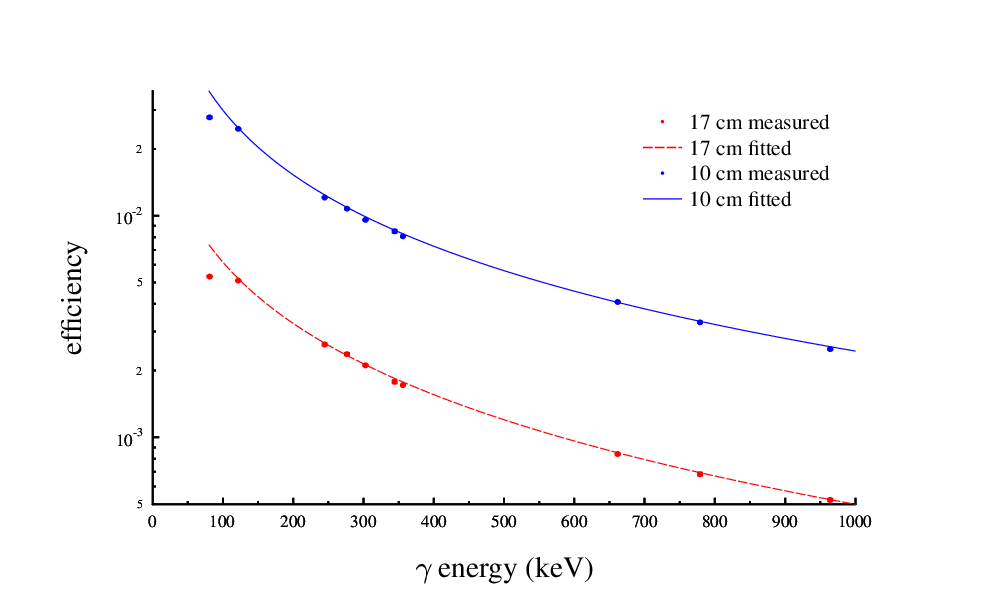}
\caption{
  \label{fig:efficiency}
  Full-energy peak efficiency of the planar detector at two source-to-detector distances. The efficiencies measured with calibration sources as well as their fitted curves are shown. The fitted efficiencies were used at the relevant $\gamma$ energies with the exception of the 79.6 and 81 keV lines from $^{133}$Xe$^m$. At these energies the fitted curves clearly deviate from the measured point. For these transitions, the efficiency measured directly with the $^{133}$Ba source was used, which has its $\gamma$ lines at these energies. 
}
\end{figure}

The absolute detection efficiency of the detectors was measured with calibrated $^{133}$Ba, $^{137}$Cs and $^{152}$Eu standard sources. The efficiency measurements were carried out at two geometries in the case of both detectors: the distance between the source and the detector surface was 10 and 17 cm at the planar detector and 10 and 27 cm at the coaxial detector. At these relatively far distances the true coincidence summing effect is negligible (below 1\,\%), thus the multi-line calibration sources can be used to obtain a smooth energy-efficiency function. As an example, the measured and fitted efficiency of the planar detector can be seen in Fig.\,\ref{fig:efficiency}.

At the lowest measured energies, however, the small cross sections necessitated the $\gamma$ counting in close geometry, placing the targets at 1\,cm from the detector surface. At such a distance the direct measurement of the efficiency is hampered by the true coincidence summing effect. Such a direct measurement was thus not made. The two-distance method \cite{Gyurky2019} was used instead to indirectly determine the detection efficiency for the relevant $\gamma$ energies. The activity of some Te targets irradiated at higher $\alpha$ energies was measured at the 1\,cm distance as well as at larger distances where the direct efficiency measurements were implemented. Distance conversion factors were then calculated for all relevant $\gamma$ energies. These conversion factors already include the effect of summing at close distances which is thus naturally accounted for. 

As it is evident from Fig.\,\ref{fig:efficiency}, the lowest energy measured efficiency points (at 79.6 and 81.0\,keV from $^{133}$Ba) are well below the fitted curve. These energies are important for the $^{130}$Te\ran$^{133}$Xe$^g$ reaction. However, since the decay of $^{133}$Xe$^g$ involves the emission of the same $\gamma$ radiation as the $^{133}$Ba calibration source, in the case of the 79.6 and 81.0\,keV lines the directly determined efficiency value was used instead of the fitted curve. 

\begin{figure*}
\centering
\includegraphics[width=\textwidth]{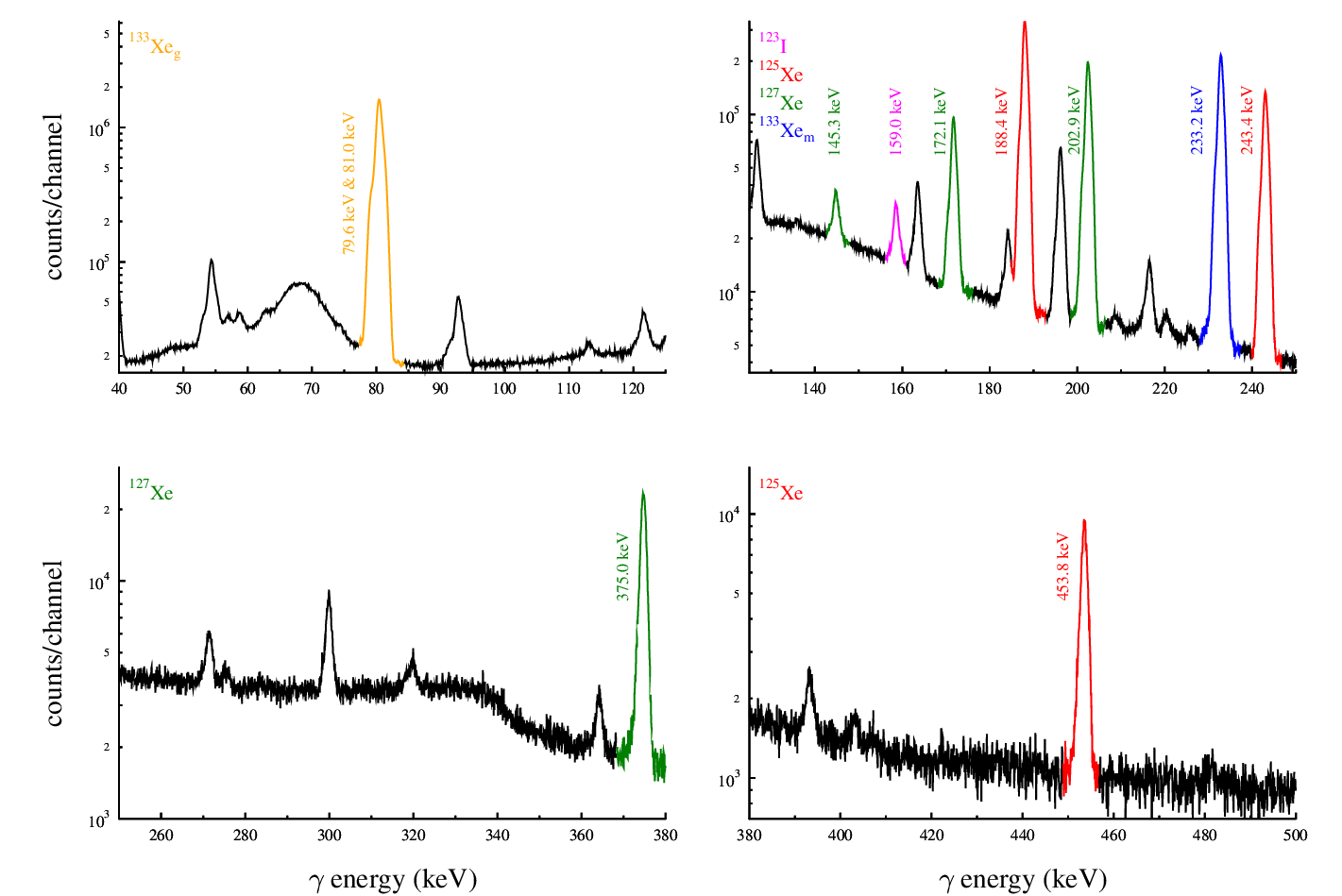}
\caption{
  \label{fig:gammaspec}
  $\gamma$ spectrum measured with the planar detector on a target irradiated with a 16\,MeV $\alpha$ beam. The spectrum is split into four panels with different vertical scales in order to see the relevant $\gamma$ peaks. The peaks used for the analyis are labeled. The other visible peaks correspond either to beam-induced or laboratory background or weaker transitions from Xe isotopes not used for the analysis.
}
\end{figure*}

Figure\,\ref{fig:gammaspec} shows a typical $\gamma$ spectrum measured with the planar detector at 10\,cm distance on the target irradiated with a 16\,MeV $\alpha$ beam. The peaks corresponding to the transitions listed in Table\,\ref{tab:decay} are indicated. 

\subsection{Results}
\label{sec:results}

Tables \ref{tab:120res}, \ref{tab:122res}, \ref{tab:124res} and \ref{tab:130res} show the obtained cross section results for the four studied reactions, respectively. Owing to the very small natural abundance of $^{120}$Te, the yield of $^{120}$Te\ran$^{123}$Xe was too small at the lowest measured energy of $E_\alpha$\,=\,10\,MeV. The cross section of all other studied reaction channels was measured in the 10-17\,MeV energy range. 

The first column of the Tables shows the primary beam energies as provided by the cyclotron. These energy values have an uncertainty of 0.3\,\% from the accelerator energy calibration. The only exception is the lowest energy measurement where a degrader foil was applied, as discussed in Sec.\,\ref{sec:irrad}. Here the first column gives the mean $\alpha$ beam energy after passing through the degrader foil. The initial $\alpha$ energy in this case was 11\,MeV. 

The second column contains the effective energies in the center-of-mass frame, which take into account the energy loss of the beam in the target layers. This energy loss, obtained with the SRIM code using the known target thicknesses, is between 80 and 125 keV. In such an energy interval the \ran\ cross sections change only moderately, but this change is not known a priory and varies from channel to channel. Therefore, the effective energy is simply assigned to the middle of the target where half of the energy loss is reached. The quoted uncertainty of the effective energies correspond to the whole $\alpha$-energy range covered by the target and also takes into account the uncertainties of the primary beam energy, the target thickness and the stopping power.

The cross section values for the measured reactions are listed in the tables with their total uncertainties. These uncertainties are obtained as the quadratic sum of the following components: statistical uncertainty from the $\gamma$-spectrum analysis (always $<$\,35\%, in most cases $<$\,5\%) and systematic uncertainties from beam charge integration (3\,\%), target thickness (4\,\%), detection efficiency including geometry conversion factors (3-5\,\%) and decay parameters ($<$\,6\%). Owing to the recent high-precision measurement \cite{Murugan2020}, the uncertainty of the natural Te isotopic abundances are well below 1\,\% (including the previously poorly known $^{120}$Te \cite{IUPAC2016}) and are therefore negligible. 
 
The rightmost column of the tables shows the astrophysical S-factor values calculated from the cross sections \cite{Iliadis2015}. These S-factors will be used in the figures in the next section dealing with the analysis of the results. The quoted S-factor uncertainties do not include the energy uncertainties. 

Since the cross sections were measured with the activation method and natural isotopic composition targets were used, different reaction channels leading to the same produced isotope cannot be distinguished. In the studied energy range such disturbing reaction channels are the following: 
\begin{itemize}
\item
$^{123}$Te($\alpha$,2n)$^{125}$Xe, leading to the same reaction product as $^{122}$Te($\alpha$,n)$^{125}$Xe. The threshold of this reaction is at 16.08\,MeV. Therefore, it can contribute to the measured $^{125}$Xe decay yield at the highest measured energy of $E_\alpha$\,=\,17\,MeV. The last row in Table\,\ref{tab:122res} typeset in italic shows thus not the pure $^{122}$Te($\alpha$,n)$^{125}$Xe cross section, but the sum of the two cross sections weighted by the isotopic ratios: $\sigma$({\footnotesize $^{122}$Te($\alpha$,n)$^{125}$Xe}) + 0.884/2.529$\cdot\sigma$({\footnotesize $^{123}$Te($\alpha$,2n)$^{125}$Xe})
\item
$^{125}$Te($\alpha$,2n)$^{127}$Xe, leading to the same reaction product as $^{124}$Te($\alpha$,n)$^{127}$Xe. The threshold of this reaction is at 14.77\,MeV. Therefore, it can contribute to the measured $^{127}$Xe decay yield at the three highest measured energies of $E_\alpha$\,=\,15, 16 and 17\,MeV. The last three rows in Table\,\ref{tab:124res} typeset in italic show thus not the pure $^{124}$Te($\alpha$,n)$^{127}$Xe cross section, but the sum of the two cross sections weighted by the isotopic ratios: $\sigma$({\footnotesize $^{124}$Te($\alpha$,n)$^{127}$Xe}) + 7.049/4.716$\cdot\sigma$({\footnotesize $^{125}$Te($\alpha$,2n)$^{127}$Xe})
\item
\teiii \rag \xetwovii , leading to the same reaction product as
$^{124}$Te($\alpha$,n)$^{127}$Xe. In general, \rag\ cross sections are much
lower than \ran\ cross sections, and the natural abundance of \teiii\ is about
a factor of five smaller than the abundance of \teiv . Using TALYS default
parameters for the $\gamma$-strength function and the level density, the
contribution of the \teiii \rag \xetwovii\ reaction to the \xetwovii\ yield is
about 0.1\% at the lowest energy of the present study and even lower at higher
energies. We have investigated the range of calculated \rag\ cross sections
for all $\gamma$-strength functions and level densities which are available in
TALYS. For practically all combinations the yield from \teiii \rag \xetwovii\
remains far below one per cent, and even using the highest calculated \teiii
\rag \xetwovii\ cross section contributes only by about 2.5 per cent at the
lowest energy and less than 0.2 per cent at the highest energy of the present
study. Thus, the contribution of the \teiii \rag \xetwovii\ reaction 
to the \xetwovii\ yield is by far within the experimental uncertainties and
can be safely neglected.
\end{itemize}

The effect of the \rann\ reaction channels will be taken into account in the
theoretical analysis in the next section.

\begin{table}
\caption{\label{tab:120res} Experimental cross section and S-factor results of the $^{120}$Te\ran$^{123}$Xe reaction. See text for details.}
\begin{ruledtabular}
\begin{tabular}{lrclrclrcl}
%\hline
$E_\alpha$  & \multicolumn{3}{c}{$E^{\rm eff}_{\rm c.m.}$} & \multicolumn{3}{c}{Cross section} & \multicolumn{3}{c}{S-factor} \\
MeV         & \multicolumn{3}{c}{MeV}					         &	\multicolumn{3}{c}{mbarn}			 &  \multicolumn{3}{c}{10$^{25}$keV$\cdot$barn}\\
\hline
11.0	&	10.586	&	$\pm$	&	0.072	&	0.336	&	$\pm$	&	0.134	&	385	&	$\pm$	&	154	\\
12.0	&	11.557	&	$\pm$	&	0.070	&	1.80	&	$\pm$	&	0.23	&	156	&	$\pm$	&	20	\\
13.0	&	12.525	&	$\pm$	&	0.071	&	13.1	&	$\pm$	&	1.0	&	117	&	$\pm$	&	9	\\
14.0	&	13.497	&	$\pm$	&	0.069	&	49.9	&	$\pm$	&	3.9	&	58.9	&	$\pm$	&	4.7	\\
15.0	&	14.468	&	$\pm$	&	0.068	&	142	&	$\pm$	&	12	&	27.3	&	$\pm$	&	2.3	\\
16.0	&	15.438	&	$\pm$	&	0.068	&	260	&	$\pm$	&	21	&	9.8	&	$\pm$	&	0.8	\\
17.0	&	16.410	&	$\pm$	&	0.066	&	345	&	$\pm$	&	27	&	2.93	&	$\pm$	&	0.23	\\
\end{tabular}
\end{ruledtabular}
\vspace{-4mm}
\end{table}

\begin{table}
\caption{\label{tab:122res} Experimental cross section and S-factor results of the $^{122}$Te\ran$^{125}$Xe reaction. The cross section in the last row (typeset in italic) is not purely from the $^{122}$Te\ran$^{125}$Xe reaction, but has contribution from the $^{123}$Te($\alpha$,2n)$^{125}$Xe reaction. See text for details.}
\begin{ruledtabular}
\begin{tabular}{lrclrclrcl}
%\hline
$E_\alpha$  & \multicolumn{3}{c}{$E^{\rm eff}_{\rm c.m.}$} & \multicolumn{3}{c}{Cross section} & \multicolumn{3}{c}{S-factor} \\
MeV         & \multicolumn{3}{c}{MeV}					         &	\multicolumn{3}{c}{mbarn}			 &  \multicolumn{3}{c}{10$^{25}$keV$\cdot$barn}\\
\hline
10.0	&	9.612	&	$\pm$	&	0.082	&	0.0172	&	$\pm$	&	0.0033	&	396	&	$\pm$	&	77	\\
11.0	&	10.591	&	$\pm$	&	0.072	&	0.319	&	$\pm$	&	0.030	&	367	&	$\pm$	&	35	\\
12.0	&	11.563	&	$\pm$	&	0.070	&	2.31	&	$\pm$	&	0.18	&	200	&	$\pm$	&	16	\\
13.0	&	12.531	&	$\pm$	&	0.071	&	14.0	&	$\pm$	&	1.1	&	125	&	$\pm$	&	10	\\
14.0	&	13.504	&	$\pm$	&	0.069	&	58.6	&	$\pm$	&	4.5	&	69.3	&	$\pm$	&	5.3	\\
15.0	&	14.476	&	$\pm$	&	0.068	&	149	&	$\pm$	&	11	&	28.8	&	$\pm$	&	2.2	\\
16.0	&	15.446	&	$\pm$	&	0.068	&	271	&	$\pm$	&	21	&	10.2	&	$\pm$	&	0.8	\\
\textit{17.0}	&	\textit{16.419}	&	$\pm$	&	\textit{0.066}	&	\textit{380}	&	$\pm$	&	\textit{30}	&	\textit{3.23} &	$\pm$	&	\textit{0.25}	\\
\end{tabular}
\end{ruledtabular}
\vspace{-4mm}
\end{table}

\begin{table}
\caption{\label{tab:124res} Experimental cross section and S-factor results of the $^{124}$Te\ran$^{127}$Xe reaction. The cross sections in the last three rows (typeset in italic) are not purely from the $^{124}$Te\ran$^{127}$Xe reaction, but have contribution from the $^{125}$Te($\alpha$,2n)$^{127}$Xe reaction. See text for details.}
\begin{ruledtabular}
\begin{tabular}{lrclrclrcl}
%\hline
$E_\alpha$  & \multicolumn{3}{c}{$E^{\rm eff}_{\rm c.m.}$} & \multicolumn{3}{c}{Cross section} & \multicolumn{3}{c}{S-factor} \\
MeV         & \multicolumn{3}{c}{MeV}					         &	\multicolumn{3}{c}{mbarn}			 &  \multicolumn{3}{c}{10$^{25}$keV$\cdot$barn}\\
\hline
10.0	&	9.617	&	$\pm$	&	0.082	&	0.0300	&	$\pm$	&	0.0063	&	689	&	$\pm$	&	144	\\
11.0	&	10.597	&	$\pm$	&	0.072	&	0.369	&	$\pm$	&	0.038	&	424	&	$\pm$	&	44	\\
12.0	&	11.569	&	$\pm$	&	0.070	&	2.45	&	$\pm$	&	0.22	&	212	&	$\pm$	&	19	\\
13.0	&	12.538	&	$\pm$	&	0.071	&	15.5	&	$\pm$	&	1.2	&	139	&	$\pm$	&	11	\\
14.0	&	13.511	&	$\pm$	&	0.069	&	60.5	&	$\pm$	&	4.9	&	71.6	&	$\pm$	&	5.8	\\
\textit{15.0}	&	\textit{14.483}	&	$\pm$	&	\textit{0.068}	&	\textit{159}	&	$\pm$	&	\textit{13}	&	\textit{30.7}	&	$\pm$	&	\textit{2.5}	\\
\textit{16.0}	&	\textit{15.454}	&	$\pm$	&	\textit{0.068}	&	\textit{365}	&	$\pm$	&	\textit{29}	&	\textit{13.7}	&	$\pm$	&	\textit{1.1}	\\
\textit{17.0}	&	\textit{16.428}	&	$\pm$	&	\textit{0.066}	&	\textit{619}	&	$\pm$	&	\textit{50}	&	\textit{5.25}	&	$\pm$	&	\textit{0.42}	\\
\end{tabular}
\end{ruledtabular}
\vspace{-4mm}
\end{table}

\begin{table*}
\caption{\label{tab:130res} Experimental cross section and S-factor results of the $^{130}$Te\ran$^{133}$Xe reaction. Partial cross sections leading to the ground and isomeric states of $^{133}$Xe are listed separately as well as their sum as the total cross section. The quoted S-factors correspond to the total cross section. See text for details.}
\begin{ruledtabular}
\begin{tabular}{lrclrclrclrclrcl}
%\hline
$E_\alpha$  & \multicolumn{3}{c}{$E^{\rm eff}_{\rm c.m.}$} & \multicolumn{9}{c}{Cross section [mbarn]}		 & \multicolumn{3}{c}{S-factor} \\
\cline{5-13}
MeV         & \multicolumn{3}{c}{MeV}					         &	\multicolumn{3}{c}{Ground state} & \multicolumn{3}{c}{Isomeric state} & \multicolumn{3}{c}{Total}	 &  \multicolumn{3}{c}{10$^{25}$keV$\cdot$barn}\\
\hline
10.0	&	9.631	&	$\pm$	&	0.082	&	0.0263	&	$\pm$	&	0.0030	&	0.0127	&	$\pm$	&	0.0023	&	0.0390	&	$\pm$	&	0.0042	&	899.1	&	$\pm$	&	97.8	\\
11.0	&	10.612	&	$\pm$	&	0.072	&	0.315	&	$\pm$	&	0.030	&	0.143	&	$\pm$	&	0.016	&	0.457	&	$\pm$	&	0.041	&	526.4	&	$\pm$	&	47.1	\\
12.0	&	11.585	&	$\pm$	&	0.070	&	2.42	&	$\pm$	&	0.19	&	1.127	&	$\pm$	&	0.089	&	3.54	&	$\pm$	&	0.27	&	307.6	&	$\pm$	&	23.8	\\
13.0	&	12.556	&	$\pm$	&	0.072	&	10.5	&	$\pm$	&	0.8	&	6.14	&	$\pm$	&	0.49	&	16.6	&	$\pm$	&	1.3	&	149.5	&	$\pm$	&	11.7	\\
14.0	&	13.531	&	$\pm$	&	0.069	&	26.9	&	$\pm$	&	2.1	&	22.0	&	$\pm$	&	1.7	&	48.9	&	$\pm$	&	3.8	&	57.9	&	$\pm$	&	4.5	\\
15.0	&	14.504	&	$\pm$	&	0.068	&	36.4	&	$\pm$	&	2.8	&	43.3	&	$\pm$	&	3.3	&	79.7	&	$\pm$	&	6.1	&	15.4	&	$\pm$	&	1.2	\\
16.0	&	15.477	&	$\pm$	&	0.068	&	35.6	&	$\pm$	&	2.8	&	56.0	&	$\pm$	&	4.2	&	91.6	&	$\pm$	&	7.1	&	3.46	&	$\pm$	&	0.27	\\
17.0	&	16.451	&	$\pm$	&	0.066	&	28.5	&	$\pm$	&	2.3	&	49.5	&	$\pm$	&	3.8	&	78.0	&	$\pm$	&	6.0	&	0.66	&	$\pm$	&	0.05	\\
\end{tabular}
\end{ruledtabular}
\vspace{-4mm}
\end{table*}

\section{Theoretical analysis}
\label{sec:theo}
\subsection{General remarks}
\label{sec:gen}
A major goal of the present work is the study of \al -nucleus optical model
potentials (AOMPs). As already pointed out in the introduction, angular
distributions of elastic scattering are the basic building block for the
determination of the AOMP. More than fifty years ago, McFadden and Satchler
(MCF) have determined a simple 4-parameter AOMP \cite{MCF1966}: The depths
$V_0$ and $W_0$, the radius $R_0$ (with $R = R_0 \times A_T^{1/3}$) and the
diffuseness $a$ of the real and imaginary part of a volume Woods-Saxon (WS)
potential were adjusted to fit a wide range of elastic scattering angular
distributions around 25 MeV. Later, additional cross section data from \al
-induced reactions like \rag , \ran , and \rap\ were included in the analysis,
and the real part of the potential was replaced by a folding potential
\cite{Demetriou_NPA2002_aomp}. Three different versions of this Demetriou
{\it et al.}\ (in the following: DEM) AOMP were provided which use different
shapes of the imaginary part and a dispersive coupling of the real and
imaginary parts in the third version (DEM3). Avrigeanu {\it et al.}\ (in the following: AVR)
\cite{Avrigeanu_NPA2003_aomp,Avrigeanu_PRC2014_aomp} went back to an WS-shaped
AOMP where many parameters were adjusted to match a wide range of available
\al -induced \raX\ data. The benefit of the DEM and AVR approaches is a better
reproduction of the \raX\ data when compared to the MCF approach, in
particular towards lower energies below the Coulomb barrier which is the
astrophysically relevant energy region. Disadvantages are the increasing
number of free parameters and a strong sensitivity to the chosen \raX\
data. In detail, the latter sensitivity to a particular data set for
\tetwonull \ran \xetwoiii\ by Palumbo {\it et al.} \cite{Palumbo_PRC2012_a-n}
will be illustrated below. 

In the recent years, it was noticed that the calculation of \raX\ reaction
cross sections at very low energies depends very sensitively on the strength
of the imaginary part of the AOMP at large radii (far beyond the colliding
nuclei) which is not at all constrained by elastic scattering and only poorly
constrained by \raX\ data. Therefore, a new AOMP was introduced which is based
purely on the barrier transmission model \cite{Mohr_PRL2020}. The
underlying real part of the AOMP was taken from the folding approach. The few
parameters were adjusted -- in a similar way as in the MCF potential -- to
elastic scattering angular distributions at low energies
\cite{Mohr_ADNDT2013_atomki-v1}. As the few parameters of this so-called
Atomki-V2 AOMP are fully fixed to elastic scattering, the Atomki-V2 AOMP
allows the robust prediction of \raX\ cross sections without further parameter
adjustment. It has been found in \cite{Mohr_PRL2020} that these predictions from
the Atomki-V2 AOMP are robust within less than a factor of two for all
\al-induced \raX\ cross sections of heavy target nuclei at low energies. This
is further confirmed by recent measurements
\cite{Kiss_APJ2021_zr96an,Szegedi2021,Ong_PRC2022_mo100an,Gyurky2023}. A
database of astrophysical \raX\ reaction rates from the Atomki-V2 AOMP is
already provided in \cite{Mohr_ADNDT2021}.

\subsection{Statistical model and ingredients}
\label{sec:stat}
The following calculations of the \ran\ reaction cross sections are based on
the statistical model \cite{Hauser1952}. In a schematic notation,
the cross section of an \al -induced \raX\ reaction is given by
\begin{equation}
\sigma(\alpha,X) \sim \frac{T_{\alpha,0} T_X}{\sum_{T_i}} = T_{\alpha,0}
  \times b_X
\label{eq:SM}
\end{equation}
with the transmission coefficients $T_{\alpha,0}$ of the incoming \al
-particle, $T_i$ for the outgoing particles ($i = \gamma, p, n, \alpha, 2n$,
etc.), and the branching ratio $b_X = T_X / \sum_i T_i$ for the branching into
the $X$ channel. Usually, the transmissions $T_i$ are calculated from optical
model potentials for the particle channels and from the \g -ray strength
function for the \rag\ capture channel. All $T_i$ depend implicitly on the
chosen level densities for the residual nuclei. For further details, see
e.g.~\cite{Rauscher_ADNDT2000_rates,Rauscher2011}. The TALYS code
\cite{TALYS-general,TALYS2023} was used for the following calculations.

For heavy target nuclei, there are common properties for the branching
$b_X = T_X / \sum_i T_i$ in Eq.\,(\ref{eq:SM}). At low energies below the \ran\
threshold, the transmissions for charged particles are strongly suppressed by
the Coulomb barrier, and thus the dominating transmission is $T_\gamma$ for
the photon channel. This leads to $b_\gamma \approx 1$ below the \ran\
threshold. Above the \ran\ threshold, neutron emission is dominating which
leads to $b_n \approx 1$ already close above the \ran\ threshold. At high
energies above the \rann\ threshold, the \ran\ and \rann\ channels compete,
leading to $b_n + b_{nn} \approx 1$. As soon as $b_X \approx 1$ is found for a
particular channel, the corresponding \raX\ cross section depends essentially
only on $T_{\alpha,0}$ which in turn only depends on the chosen AOMP but not
on the other ingredients of the statistical model. Hence we focus on the role
of the AOMP in the subsequent analysis whereas the other ingredients of the
statistical model play only a very minor role. This discussion is extended at
the end of Sec.~\ref{sec:technical}.

\begin{figure}
\centering
\includegraphics[width=\columnwidth]{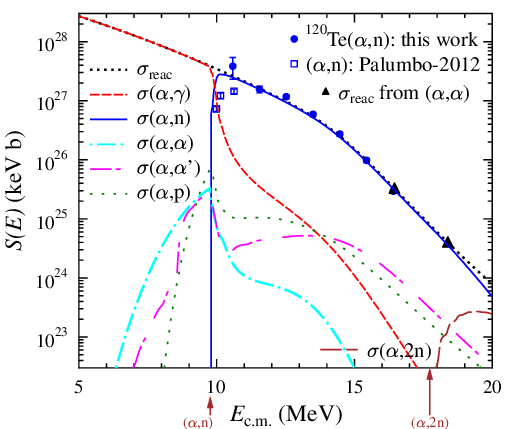}
\caption{
  \label{fig:te120a_branching}
  Decomposition of the total reaction cross section \stot\ (black dotted) of
  \tetwonull\ + \al\ into the contributions of the different \raX\ channels. The
  \rag\ channel (red dashed) is dominating below the \ran\ threshold. The
  \ran\ channel (blue) becomes dominating within about 1 MeV above the \ran\
  threshold. Only at high energies around 20 MeV, the contribution of the
  \rann\ channel (brown long-dashed) contributes significantly. The
  contributions of the other open channels remain negligible for all energies
  under study. The vertical arrows show the \ran\ and \rann\ thresholds. All
  cross sections have been converted to astrophysical \sfact s.
}
\end{figure}

The above considerations on the branching $b_X$ are visualized for
\tetwonull\ + \al\ in Fig.\,\ref{fig:te120a_branching}. For better
visibility, all cross sections are converted to astrophysical \sfact s which
show only a moderate energy dependence (compared to the steeper energy
dependence of the underlying cross sections).

\subsection{Additional data from elastic scattering}
\label{sec:elast}
\begin{figure}
\centering
\includegraphics[width=\columnwidth]{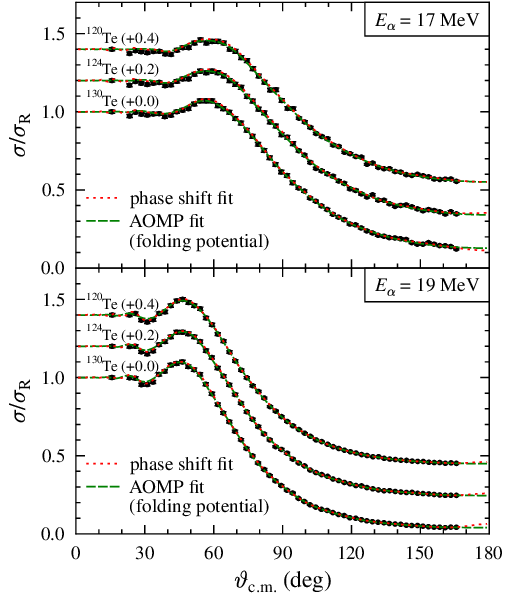}
\caption{
  \label{fig:te_elast}
  Elastic scattering angular distributions for \tetwonull \raa \tetwonull ,
  \teiv \raa \teiv , and \tethreenull \raa \tethreenull\ at 17 MeV (upper
  part) and 19 MeV (lower part) : The experimental data of
  \cite{Palumbo_PRC2012_a-a} are well reproduced by phase shift fits (red
  dotted) and AOMP fits (green dashed). Both lines overlap almost fully; the
  differences are hardly visible. The resulting total reaction cross sections
  \stot\ are listed in Table \ref{tab:elast}. Further explanations see text.
}
\end{figure}

As pointed out above, the experimental \ran\ cross sections provide an
excellent constraint for the AOMP because the calculated \ran\ cross sections
are mainly sensitive to the AOMP only. At higher energies, the total reaction
cross section \stot\ of \al -induced reactions can also be used to test the
AOMP. \stot\ is derived from the analysis of the elastic scattering angular
distributions. Full angular distributions at low energies have been measured
by Palumbo {\it et al.}\ \cite{Palumbo_PRC2012_a-a}, but unfortunately \stot\
was not determined in \cite{Palumbo_PRC2012_a-a}. Hence we have re-analyzed
the data of \cite{Palumbo_PRC2012_a-a} for \tetwonull , \teiv , and
\tethreenull\ below 20 MeV using either phase shift fits or AOMP
fits. Fig.~\ref{fig:te_elast} shows that the \raa\ angular distributions of
Palumbo {\it et al.}\ \cite{Palumbo_PRC2012_a-a} at 17 MeV and at 19 MeV are
very well reproduced by both approaches. The phase shift fits (red dotted
lines in Fig.~\ref{fig:te_elast}) achieve $\chi^2/F \lesssim 1.0$, and the
AOMP fits (based on a folding potential in the real part and a surface
Woods-Saxon imaginary part) show slightly higher $\chi^2/F \approx 1.0 - 1.5$
(green dashed lines in Fig.~\ref{fig:te_elast}).

The resulting \stot\ are calculated from the reflexion coefficients $\eta_L$
of the phase shift fits and of the AOMP fits. The \stot\ from both approaches
agree within a few per cent in all cases under study. The average of both
approaches is finally given as \stot\ in Table \ref{tab:elast}. The analysis
benefits from the fact that the angular distributions in
\cite{Palumbo_PRC2012_a-a} cover the almost full angular range with small
uncertainties.

For comparison, we include another data point for \stot\ of \teiv \raa \teiv\
from the data of \cite{SalemVasconcelos_NPA1979} at 19.3 MeV. A calculation of
the total reaction cross section \stot\ from the optical model parameters in
Table 1 of \cite{SalemVasconcelos_NPA1979} results in \stot\ $= 969$ mb, i.e.\
significantly above the corresponding \stot\ derived from the Palumbo {\it et
  al.}\ data \cite{Palumbo_PRC2012_a-a}. However, the calculated angular
distribution from these parameters in Table 1 of \cite{SalemVasconcelos_NPA1979}
does not match the experimental data in their Fig.~2. A much better agreement
is obtained when the radius parameter $r_0$ of 1.395 fm is replaced by 1.24
fm. The latter value is given in several other lines in Table 1 of
\cite{SalemVasconcelos_NPA1979}; this points to a simple typo in Table 1. Using
$r_0 = 1.24$ fm, we find \stot\ $= 660$ mb, consistent with the results by
Palumbo {\it et al.}; hence, this value is listed in Table \ref{tab:elast}. An
attempt to re-fit the angular distribution of \cite{SalemVasconcelos_NPA1979}
is hampered by the fact that the data at EXFOR had to be re-digitized from
Fig.~2 which results in additional uncertainties. Furthermore, the data in
Fig.~2 of \cite{SalemVasconcelos_NPA1979} do not show experimental
uncertainties. A re-fit of the EXFOR data, assuming a constant uncertainty of
5\% for all data points, leads to slightly higher \stot\ between about 700 mb
and 770 mb. The larger scatter of the derived \stot\ in the fits results from
the smaller angular range and the lower number of experimental data points
which do not allow a phase shift analysis. Overall, despite the above
uncertainties under discussion, the data at 19.3 MeV by
\cite{SalemVasconcelos_NPA1979} confirm the consistency of the \stot\ from the
experimental scattering data of \cite{Palumbo_PRC2012_a-a} within the given
uncertainties.
\begin{table}[htb]
  \caption{
    \label{tab:elast}
Total reaction cross sections \stot\ for various tellurium isotopes, derived
from elastic scattering \cite{Palumbo_PRC2012_a-a}. One independent data point
for \teiv\ at 19.3 MeV \cite{SalemVasconcelos_NPA1979} confirms the present
results.
}
\begin{center}
\begin{tabular}{ccr@{$\pm$}l}
\hline
\multicolumn{1}{c}{isotope}
& \multicolumn{1}{c}{$E_\alpha$ (MeV)} 
& \multicolumn{2}{c}{\stot\ (mb)} \\
\hline
  \tetwonull   & 17.0 & 397 & 20 \\
  \tetwonull   & 19.0 & 648 & 32 \\
  \teiv        & 17.0 & 407 & 20 \\
  \teiv        & 19.0 & 673 & 34 \\
  \teiv        & 19.3 & \multicolumn{2}{c}{660 \cite{SalemVasconcelos_NPA1979}} \\
  \tethreenull & 17.0 & 444 & 22 \\
  \tethreenull & 19.0 & 711 & 36 \\  
\hline
\end{tabular}
\end{center}
\end{table}

Note that the data points for the total reaction cross section \stot\ from
elastic scattering at 17 MeV by Palumbo {\it et al.}\
\cite{Palumbo_PRC2012_a-a} are shown in all figures for \tetwonull , \teiv ,
and \tethreenull\ as black triangles, whereas the 19 MeV data are above the
energy range of the present \ran\ data and thus outside the chosen scale of
some figures. For clarification we point out that the three data points for
the \tetwonull \ran \xetwoiii\ reaction by Palumbo {\it et al.}\
\cite{Palumbo_PRC2012_a-n} are shown in the respective figures for \tetwonull\
by blue squares.

\subsection{Further technical remarks}
\label{sec:technical}
\begin{figure}
\centering
\includegraphics[width=\columnwidth]{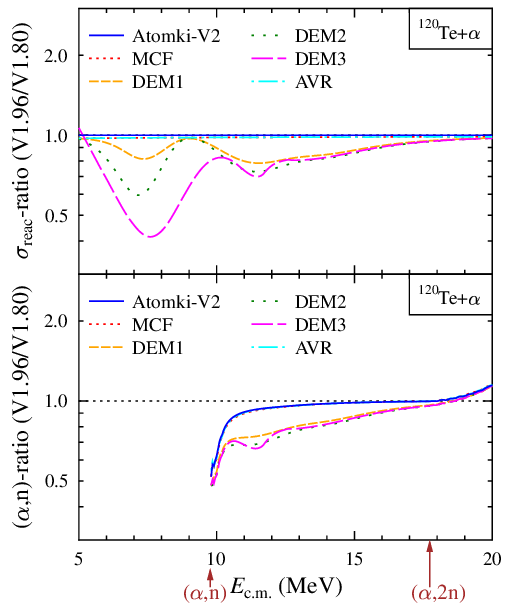}
\caption{
  \label{fig:talys_ratio}
  Comparison of the TALYS versions 1.80 and 1.96 for the total reaction cross
  section \stot\ (upper) and the \ran\ cross section (lower) for the example
  of \tetwonull . Further discussion see text.
}
\end{figure}

The calculations in the present study are mainly done with version 1.80 of
TALYS \cite{TALYS-general,TALYS2023}. This choice is motivated by a poorly
documented modification of the AOMPs by Demetriou {\it et al.}\ 
\cite{Demetriou_NPA2002_aomp} in later versions. For explanation, we present
the ratio between TALYS-1.80 and TALYS-1.96 for the calculated total reaction
cross section \stot\ and for the \ran\ cross section in
Fig.\,\ref{fig:talys_ratio}.

The total reaction cross section \stot\ depends only on the AOMP, and thus the
results from different TALYS versions should be identical. As expected, the
\stot\ are indeed identical for the MCF, the AVR, and the Atomki-V2 AOMPs; see
Fig.\,\ref{fig:talys_ratio}, upper part. However, the ratio deviates
significantly from unity for the DEM1, DEM2, and DEM3 AOMPs.

It has been pointed out by the TALYS authors \cite{TALYS_Goriely2024_priv}
that a modification of the DEM1, DEM2, and DEM3 AOMPs was implemented in TALYS
version 1.96 which is based on further studies of \al -induced reaction cross
sections by P. Demetriou. Unfortunately, this modification was not yet
listed in the detailed ``Log file of changes'' at the end of the TALYS manual;
this will change in the next TALYS release. At present, the explanation can
only be found as a comment in the TALYS source code ``{\it{foldalpha.f}}'':
{\it{``csg Correction of the radius dependence of the real part after an
    analysis of the (a,g) and (a,n) data of deformed nuclei : 27/4/2018
    (Brussels) increase of rva by 3\% for deformed nuclei but only below
    typically 18 MeV''}}. The source code shows that the correction of the
radius depends on the deformation parameter $\beta_2$. In the case of
\tetwonull , the negative $\beta_2$ in TALYS leads to a decrease of the radius
of the real potential (the statement in the comment is only valid for
$\beta_2 > 0$). A reduced radius of the attractive real part of the AOMP leads
to an effective increase of the Coulomb barrier and thus to lower total
reaction cross sections \stot\ (as visible in the upper part of
Fig.\,\ref{fig:talys_ratio}). We prefer to use the original DEM1, DEM2, and
DEM3 AOMPs without the undocumented modification in TALYS-1.96, and thus we
use TALYS-1.80 in the following analysis of the \ran\ cross sections.

For completeness we note that there is another difference between TALYS-1.80
and TALYS-1.96 which becomes visible in the \ran\ cross sections. The default
\g -strength function (GSF) was changed which leads to a larger GSF in
TALYS-1.96. As a consequence, close above the \ran\ threshold the \rag\
contribution is larger which in turn leads to smaller \ran\ cross
sections. This effect is mostly visible in the first MeV above the \ran\
threshold, but remains within about 10\% at higher energies. This will be
discussed further in the analysis of the \tetwonull \ran \xetwoiii\ reaction
in Sec.\,\ref{sec:te120an}. As expected, both TALYS versions in use provide
identical results as soon as the same GSF is selected.

\begin{figure}
\centering
\includegraphics[width=\columnwidth]{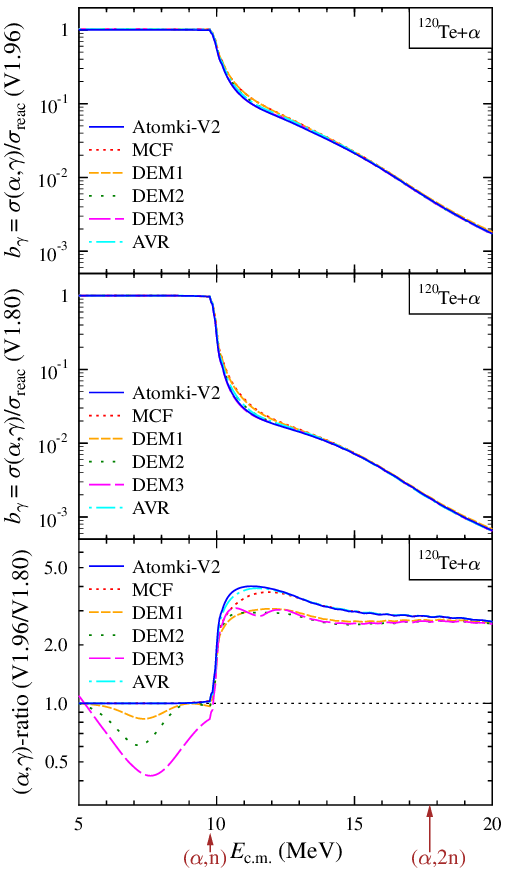}
\caption{
  \label{fig:talys_ag_ratio}
  Comparison of the TALYS versions 1.80 and 1.96 for the \tetwonull \rag
  \xetwoiv\ cross section and the relevance of the GSF: ratio of the \rag\
  cross sections of TALYS versions 1.96 and 1.80 for different AOMPs (lower
  part); ratio of the \rag\ cross section to the total cross section \stot\
  for TALYS version 1.80 (middle part); same, but for TALYS version 1.96
  (upper part). Further discussion see text.
}
\end{figure}

The results from the different default options for the GSF in the TALYS
versions 1.80 and 1.96 are shown in Fig.~\ref{fig:talys_ag_ratio} for the
\tetwonull \rag \xetwoiv\ reaction. This comparison can be used to assess the
relevance of the GSF for the present analysis.

Below the \ran\ threshold, the only open channels are \raa\ elastic and
inelastic scattering and \rag\ capture. Because of the high Coulomb barrier,
the transmission into the \al\ channel is much smaller than the transmission
into the \g\ channel: $T_\alpha \ll T_\gamma$. This leads to a branching
$b_\gamma \approx 1$ in Eq.~(\ref{eq:SM}) and thus $\sigma$\rag\ $\approx$
\stot . This holds for all AOMPs under study and for the lower GSF in TALYS
1.80 as well as for the higher GSF in TALYS 1.96 (see upper and middle parts
of Fig.~\ref{fig:talys_ag_ratio}).

Above the \ran\ threshold, the \ran\ channel dominates, and thus the branching
$b_\gamma$ in Eq.~(\ref{eq:SM}) becomes approximately
$b_\gamma \approx T_\gamma / (T_\gamma + T_n ) \approx T_\gamma / T_n$ at higher
energies. In practice, $b_\gamma$ is of the order of a few per cent close
above the \ran\ threshold and decreases to about $10^{-3}$ at higher energies
(see Fig.~\ref{fig:talys_ag_ratio}, upper and middle parts). The GSF is about
a factor of three larger in TALYS 1.96, as can be seen at the higher energies
in Fig.~\ref{fig:talys_ag_ratio}, lower part. The same factor is found for the
branching ratios $b_\gamma$. A further enhancement of the GSF (exceeding a
factor of three between the different GSFs in TALYS 1.80 and 1.96) will lead
to further increasing $b_\gamma$ and simultaneous decrease of $b_n$ and the
\ran\ cross section. However, a reduction of the GSF will have only very minor
influence because a reduction of $b_\gamma$ from e.g.\ $10^{-3}$ to $10^{-4}$
will enhance the neutron branching $b_n \approx 1 - b_\gamma$ only from 0.999
to 0.9999, will is less than 0.1 per cent.

Finally we note an interesting detail: The increased GSF in TALYS 1.96 is
responsible for the increase of the ratio of \ran\ cross sections above the
\rann\ threshold, see Fig.~\ref{fig:talys_ratio}, lower part. The increased
GSF favors the \g -decay of highly excited \xetwoiii\ residual nuclei and thus
reduces the probability of emission of a second neutron and the \rann\ cross
section.

\subsection{Comparison between experimental data and theory}
\label{sec:comp}
\subsubsection{ \tetwonull  \ran \xetwoiii }
\label{sec:te120an}
\begin{figure}
\centering
\includegraphics[width=\columnwidth]{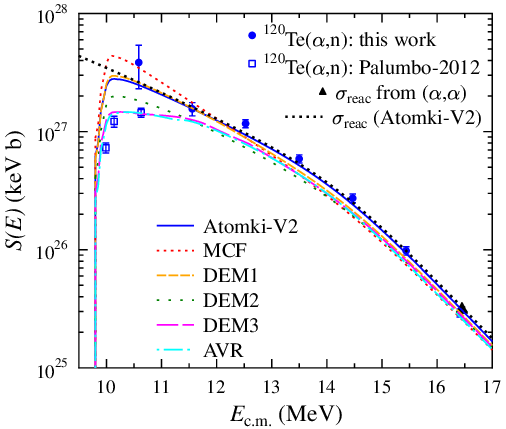}
\caption{
  \label{fig:te120an}
  Reaction cross section of the \tetwonull \ran \xetwoiii\ reaction (shown as
  astrophysical \sfact ): comparison of the experimental data to TALYS
  calculations using different AOMPs. In addition, the total reaction cross
  section \stot\ from the Atomki-V2 AOMP (black dashed) is compared to \stot\
  from elastic scattering (black triangle, see Table \ref{tab:elast}).
  The \rann\ threshold located at $E_{\rm c.m.}$\,=\,17.725\,MeV 
	is outside the energy range of the figure.
}
\end{figure}

Fig.\,\ref{fig:te120an} compares the experimental \ran\ data from this work and
from Palumbo {\it et al.}\ \cite{Palumbo_PRC2012_a-n} to TALYS calculations
using different AOMPs. It is obvious that all AOMPs except the Atomki-V2 AOMP
underestimate the experimental data above 11 MeV where the \ran\ cross
sections are mainly sensitive to the AOMP. There is excellent agreement
between the total cross section \stot\ from elastic scattering and the \ran\
cross section from this work at $E_{c.m.} \approx 16.4$ MeV (note that both
data points, black triangle and blue circle, are almost hidden behind the
various lines from the AOMP calculations).

There is some tension between the three low-energy data points by Palumbo {\it 
  et al.}\ and our lowest data point around 11 MeV. Unfortunately, there
is no overlap region where both data sets provide \ran\ cross sections with
small uncertainties. Thus, the \ran\ cross section near the \ran\
threshold remains somewhat uncertain. An improvement of the present experiment
requires enriched targets which were also used in the Palumbo {\it et al.}\
experiment.

\begin{figure}
\centering
\includegraphics[width=\columnwidth]{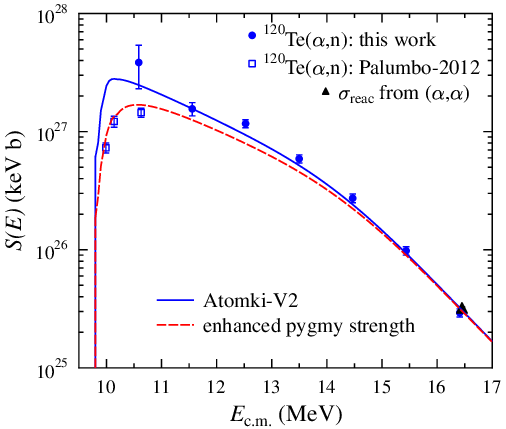}
\caption{
  \label{fig:te120an_test}
  Reaction cross section of the \tetwonull \ran \xetwoiii\ reaction (shown as
  astrophysical \sfact ) from the Atomki-V2 AOMP with the default GSF of
  TALYS-1.80 (blue) and with an additional pygmy dipole strength at low
  energies (red dashed). Further discussion see text.
}
\end{figure}

The standard calculation with the Atomki-V2 AOMP in TALYS 1.80 fits the
present data over the full energy range, favoring the high cross section at
the lowest energy of the present experiment. However, following the findings
in Sec.\,\ref{sec:technical}, the \ran\ cross section may be reduced by an
enhanced \rag\ contribution. As an example, we show in
Fig.\,\ref{fig:te120an_test} that an additional pygmy dipole strength enhances
the \rag\ cross section in such a way that the low \ran\ cross sections of
Palumbo {\it et al.}\ close above the threshold are nicely reproduced. But
such an enhanced \rag\ contribution would slightly reduce the \ran\ cross
sections up to about 15 MeV. A simultaneous fit of the low \ran\ data by
Palumbo {\it et al.}\ close above the threshold and the present data over the
full energy range is practically impossible without very special adjustments
of the GSFs and the level densities which affect the branching between the
\rag\ and \ran\ channels.

The AVR AOMP was adjusted -- among many others -- to the low \ran\ cross
sections of Palumbo {\it et al.}. Obviously, this had lead to
parameters of the AOMP which underestimate the \ran\ cross sections at higher
energies. Similar underestimations of the \ran\ cross sections are found for
the other tellurium isotopes under study in the present work (see subsequent
sections).

The DEM1, DEM2, and DEM3 AOMPs were derived before the Palumbo {\it et al.}\
experiment. Thus, there is no obvious reason for the overall underestimation
of the \ran\ cross section which holds also for the other tellurium isotopes
under study (see subsequent sections). We note that the underestimation
increases further when the undocumented modification of the DEM1, DEM2, and
DEM3 AOMPs in TALYS-1.96 is applied because this modification reduces the
\ran\ cross sections by about $10-30$\% in the enegy range of the present
study.

\subsubsection{ \teii  \ran \xetwov }
\label{sec:te122an}
\begin{figure}
\centering
\includegraphics[width=\columnwidth]{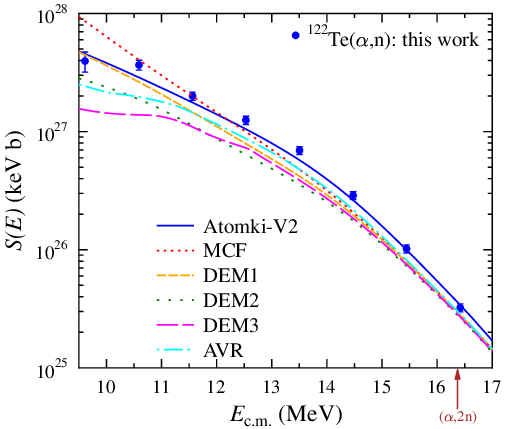}
\caption{
  \label{fig:te122an}
  Reaction cross section of the \teii \ran \xetwov\ reaction (shown as
  astrophysical \sfact ): comparison of the experimental data to TALYS
  calculations using different AOMPs. The \rann\ threshold is indicated by a
  vertical arrow.
}
\end{figure}

The results for \teii \ran \xetwov\ are shown in
Fig.\,\ref{fig:te122an}. Similar to \tetwonull \ran \xetwoiii\ in the previous
section, the new experimental data are well reproduced by the Atomki-V2
AOMP. The DEM1, DEM2, DEM3, and AVR AOMPs underestimate the new experimental
data over the full energy range. The MCF AOMP shows a much steeper energy
dependence than all other AOMPs, leading to an overestimation of the \ran\
cross section towards the lowest energies. This is a well-known finding for
the MCF AOMP for heavy target nuclei.

The new data point at the highest energy may have received a minor
contribution from the \teiii \rann \xetwov\ reaction. We estimate this
contribution with about 5\%, i.e., within the given uncertainties. No
correction is applied here. Further information on the contributions of the
\rann\ reaction can be found in the subsequent Sec.\,\ref{sec:te124an} on the
\teiv \ran \xetwovii\ reaction.

The \ran\ threshold of the \teii \ran \xetwov\ reaction is located at 8.77
MeV. Our lowest data point is located about 1 MeV above the \ran\
threshold. Unfortunately, the new experimental data do not cover the first MeV
above the \ran\ threshold which would allow a better comparison with the
\tetwonull \ran \xetwoiii\ reaction close above the threshold with the
discrepant data by Palumbo {\it et al.} and from the present study.

Again unfortunately, elastic scattering angular distributions are not
available for \teii . Thus, a comparison of the total reaction cross section
\stot\ from elastic scattering to the \ran\ data from the present experiment
is not possible. Furthermore, the EXFOR database does not show any cross
sections for the \teii \ran \xetwov\ reaction in the literature.

\subsubsection{ \teiv  \ran \xetwovii }
\label{sec:te124an}
\begin{figure}
\centering
\includegraphics[width=\columnwidth]{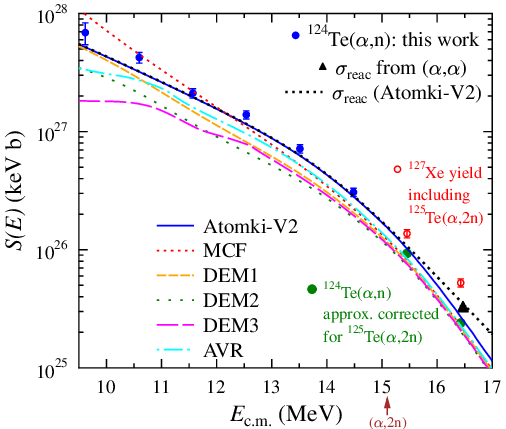}
\caption{
  \label{fig:te124an}
  Reaction cross section of the \teiv \ran \xetwovii\ reaction (shown as
  astrophysical \sfact ): comparison of the experimental data to TALYS
  calculations using different AOMPs. In addition, the total reaction cross
  section \stot\ from the Atomki-V2 AOMP (black dashed) is compared to \stot\
  from elastic scattering (black triangle, see Table \ref{tab:elast}).
  The \rann\ threshold is indicated by a vertical arrow. The contribution of
  the \tev \rann \xetwovii\ reaction is explained in the text (discussion of
  the red and green data points).
}
\end{figure}

In general, the results for the \teiv \ran \xetwovii\ reaction in
Fig.\,\ref{fig:te124an} are similar as in the two previous sections. The
Atomki-V2 AOMP provides a good description of the new experimental data
although the calculation is slightly lower than the experimental data. The
DEM1, DEM2, DEM3, and AVR AOMPs also underestimate the experimental data with
a somewhat larger deviation than the Atomki-V2 AOMP. The MCF AOMP shows a
stronger energy dependence than the other AOMPs.

Along the tellurium isotopic chain, the $Q$-value of the \ran\ reaction
becomes less negative with increasing neutron number. This holds also for the
\rann\ reactions. A first consequence is that the difference between the total
cross section \stot\ (black dashed line in Fig.\,\ref{fig:te124an}) and the
\ran\ cross section (blue line) becomes clearly visible around 16 MeV which is
about 1 MeV above the \rann\ threshold for \teiv .

A second consequence of the lowering of the $Q$-values is the fact that a
significant contribution of the \rann\ reaction on the neighboring $N+1$
isotope \tev\ has to be expected for the \xetwovii\ yield at higher
energies. As discussed in Sec.\,\ref{sec:results}, it is not possible to disentangle the contributions from the \teiv
\ran \xetwovii\ and \tev \rann \xetwovii\ reactions in the present activation
experiment. Assigning the total \xetwovii\ yield to the \teiv \ran \xetwovii\
reaction overestimates the real \ran\ cross section. This becomes nicely
visible for the two data points at 15.5 and 16.4 MeV (shown as red circles in
Fig.\,\ref{fig:te124an}). In particular, the data point at 16.4 MeV even
exceeds the total cross section \stot\ from elastic scattering which is
impossible. The \stot\ from elastic scattering can be considered as very
reliable because there is an independent confirmation for \teiv\ in
\cite{SalemVasconcelos_NPA1979} (see also Table \ref{tab:elast}).

For a rough estimate, we have calculated the expected yields for the \teiv
\ran \xetwovii\ and \tev \rann \xetwovii\ reactions in a natural target using
TALYS-1.80 in combination with the Atomki-V2 AOMP which does a good job in
general for the tellurium isotopes. This leads to a correction factor of
1/1.46 (1/2.19) at 15.5 (16.4) MeV. The correction is less than 1\% for the
next data point at $E_{\rm{c.m.}} \approx 14.5$ MeV. The resulting cross
sections for the \teiv \ran \xetwovii\ reaction (after correction for \tev
\rann \xetwovii ) are shown as full green circles in 
Fig.\,\ref{fig:te124an}. It is difficult to estimate the uncertainty of the
correction. From the results for \tethreenull\ (see next
Sec.\,\ref{sec:te130an}) one may conclude that such a theoretical correction is
slightly too strong because the \tethreenull \rann \xetwoviii\ cross section
is slightly overestimated. Nevertheless, the corrected data point at 16.4 MeV
is located below the total cross section \stot\ from elastic scattering and
thus not in contradiction to \stot .

\subsubsection{ \tethreenull  \ran \xethreeiii }
\label{sec:te130an}
\begin{figure}
\centering
\includegraphics[width=\columnwidth]{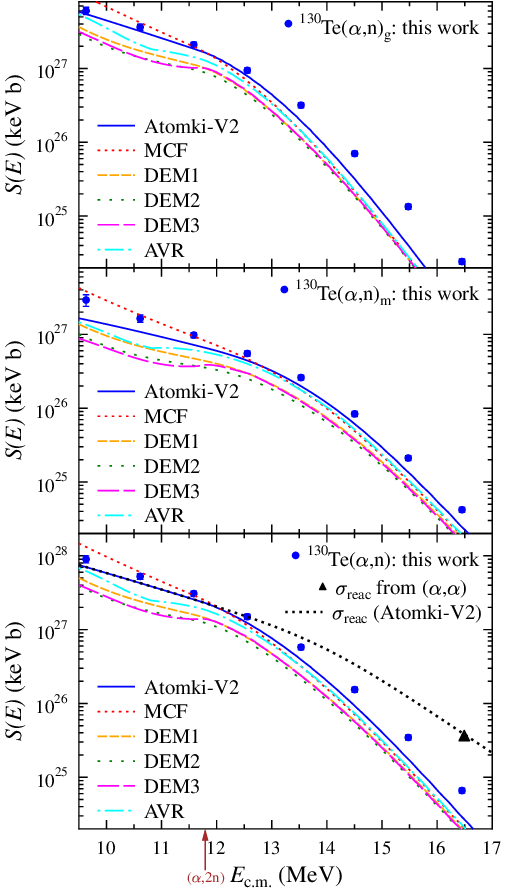}
\caption{
  \label{fig:te130an}
  Reaction cross section of the \tethreenull \ran \xethreeiii\ reaction (shown as
  astrophysical \sfact ): comparison of the experimental data to TALYS
  calculations using different AOMPs. In addition, the total reaction cross
  section \stot\ from the Atomki-V2 AOMP (black dashed) is compared to \stot\
  from elastic scattering (black triangle, see Table \ref{tab:elast}). The
  lower (middle, upper) part shows the total \ran\ (isomer, ground state)
  cross section. The \rann\ threshold is indicated by a vertical arrow.
}
\end{figure}

The results for \tethreenull \ran \xethreeiii\ are shown in
Fig.\,\ref{fig:te130an}. The first excited state in \xethreeiii\ at $E^\ast =
233$ keV with $J^\pi = 11/2^-$ is a low-lying isomer with a half-life of 2.20
days. The $3/2^+$ ground state has a longer half-life of 5.25 days. The
activation yields have been determined separately, leading to a threefold
Fig.\,\ref{fig:te130an} with the ground state contribution (upper part), the
isomer contribution (middle part), and the total \ran\ cross section from the
sum of the ground state plus isomer contributions (lower part).

Let us first focus on the total \ran\ cross section in the lower part of
Fig.\,\ref{fig:te130an}. Similar to all previous cases, it is found that the
DEM1, DEM2, DEM3, and AVR AOMPs show slightly lower cross sections than the
Atomki-V2 AOMP. However, contrary to the previous cases, also the Atomki-V2
AOMP underestimates the experimental \ran\ data significantly. At first view,
this might be misunderstood as a failure of all AOMPs. But this is not the
case. The total cross section \stot\ from elastic scattering at 16.5 MeV
clearly shows that the Atomki-V2 AOMP predicts the total cross section
correctly. Thus, the deviation between the calculated and experimental \ran\
cross sections is not related to the AOMP, but instead indicates an
overestimation of the \rann\ cross section and related underestimation of the
\ran\ cross section. Note that the summed branchings of the \ran\ and \rann\
channels are $b_n + b_{2n} \approx 1$; the contributions from other channels
are practically negligible above the \ran\ threshold (see also the discussion
of Fig.\,\ref{fig:te120a_branching} for the example of \tetwonull ). 

\begin{figure}
\centering
\includegraphics[width=\columnwidth]{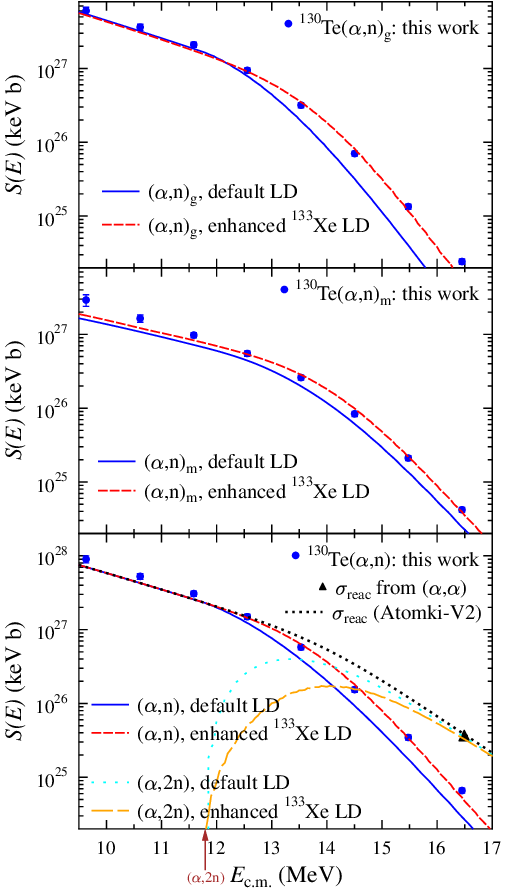}
\caption{
  \label{fig:te130an_test}
  Reaction cross section of the \tethreenull \ran \xethreeiii\ reaction (shown as
  astrophysical \sfact ): default TALYS level density (blue and light-blue)
  vs.\ enhanced level density for \xethreeiii\ (red and orange). All
  calculations are based on the Atomki-V2 AOMP. Further discussion see text.
}
\end{figure}

An artificial increase of the level density (LD) in \xethreeiii\ enhances the
contribution of the \ran\ cross section and decreases the \rann\ cross
section. This is illustrated in Fig.\,\ref{fig:te130an_test}. As the total
reaction cross section \stot\ depends only on the AOMP (but not on the chosen
LD of \xethreeiii ), \stot\ is not affected and still reproduces the data
point from elastic scattering at 16.5 MeV. At energies below the \rann\
threshold, the enhanced LD of \xethreeiii\ has also no impact because the
branching $b_n$ to the \ran\ channel is anyway close to unity. Only above the
\rann\ threshold, the enhanced LD of \xethreeiii\ increases the \ran\ cross
section and reduces the contribution of the \rann\ channel.

An even stronger increase of the LD of \xethreeiii\ could also lead to a good
description of the \ran\ cross sections using other AOMPs. However, the
underestimation of the total cross section \stot\ at 16.5 MeV from elastic
scattering would persist for the other AOMPs.

The total \tethreenull \ran \xethreeiii\ cross section is composed of the
ground state contribution and the isomer contribution. The branching between
these channels depends on \g -ray cascades in the residual nucleus \xethreeiii\
and is practically independent of the AOMP. The upper parts of
Figs.\,\ref{fig:te130an} and \ref{fig:te130an_test} show that the branching is
roughly reproduced by the calculations, with a trend to underestimate the
isomer contribution towards lower energies. This trend is slightly improved
when the enhanced level density of \xethreeiii\ is used which is required to
fit the total \ran\ cross section (see Fig.\,\ref{fig:te130an_test}).

\begin{figure}
\centering
\includegraphics[width=\columnwidth]{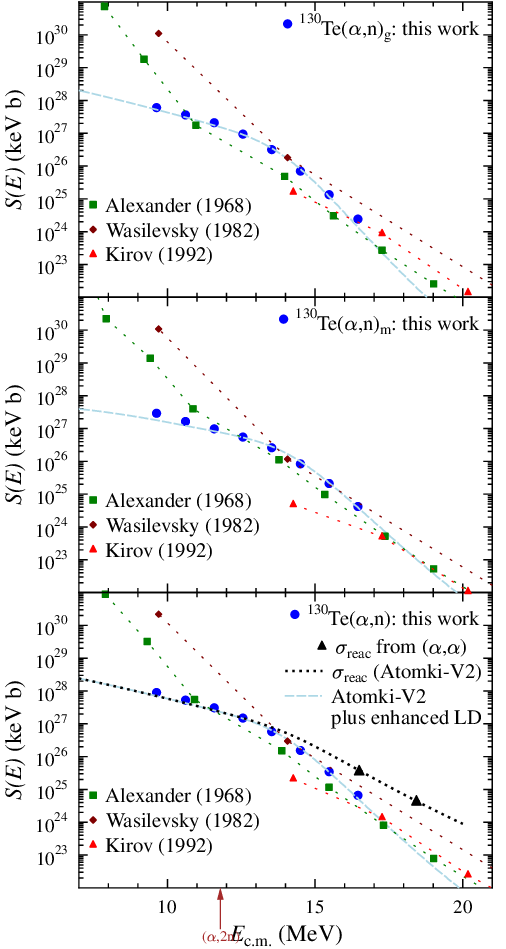}
\caption{
  \label{fig:te130an_all-exp}
  Same as Figs.\,\ref{fig:te130an} and \ref{fig:te130an_test}, but including
  experimental data from literature
  \cite{Kirov_ZPA1992,Wasilevsky_RRI1982,Alexander_NPA1968}. The data from
  each data set are connected by thin lines to guide the eye. A much larger
  scale is required to cover the wide scatter of the literature data. Further
  discussion see text.
}
\end{figure}

We have decided to include only our new data points in Figs.\,\ref{fig:te130an}
and \ref{fig:te130an_test}. Several earlier data sets
\cite{Kirov_ZPA1992,Wasilevsky_RRI1982,Alexander_NPA1968} are available for
\al -induced reactions on \tethreenull . However, all experiments used much
higher energies of the \al\ projectiles in combination with a stacked-target
technique. This typically leads to a relatively broad energy distribution at
the lowest energies with significant uncertainties from the thickness and
homogeneity of the targets and of the degrader foils. For completeness we show
the full available data in Fig.\,\ref{fig:te130an_all-exp}. Because of the huge
scatter of the experimental data, a much larger scale is required here, but
still some data are not included in the chosen scale. Most of the data from
\cite{Kirov_ZPA1992,Wasilevsky_RRI1982,Alexander_NPA1968} are located outside
of Figs.\,\ref{fig:te130an} and \ref{fig:te130an_test}. For better readability,
we show only one line with a calculation; here we choose the Atomki-V2 AOMP
in combination with the enhanced \xethreeiii\ LD from
Fig.\,\ref{fig:te130an_test} which fits our new experimental data (lightblue
dashed line). Unfortunately, it must be concluded that the low-energy data
from the stacked-foil experiments do not provide further insight because of
the huge scatter of the data, reaching several orders of magnitude in
particular towards the lowest energies, and thus by far exceeding the claimed
error bars.

For completeness we note that Kirov {\it et al.} \cite{Kirov_ZPA1992} also
provide the cross section of the \tethreenull \rap \ithreeiii\
reaction. Fig.\,\ref{fig:te130an_all-exp} shows that the \ran\ data by Kirov
{\it et al.}\ at low energies are much lower than our new data and all other
available data sets. This leads to some doubts on these data, and we do
not include the \rap\ data of Kirov {\it et al.}\ in the present
analysis. Furthermore, the \rap\ cross section depends on the combination of
many ingredients of the statistical model, but does not constrain the AOMP
directly.

\subsection{Discussion}
\label{sec:disc}
It is obvious from Figs.\,\ref{fig:te120an}, \ref{fig:te122an},
\ref{fig:te124an}, and \ref{fig:te130an} that the deviations between the new
experimental data and the calculations from various AOMPs remains limited
within a factor of about $2-3$. Huge deviations, reaching orders of magnitude,
as e.g.\ found in the pioneering experiment on $^{144}$Sm\rag $^{148}$Gd by
Somorjai {\it et al.}\ \cite{Somorjai1998} and confirmed by Scholz {\it et
  al.}\ \cite{Scholz_PRC2020_sm144ag}, are not observed for the present \ran\
data. Such huge deviations are typically found only at deep sub-barrier
energies which is below the \ran\ threshold for most heavy stable nuclei. This
will be further investigated at the end of the discussion (see also
Fig.\,\ref{fig:te120a_total_low} below).

Although the overall description of the new \ran\ data is good, there are
significant differences between the various AOMPs under study. The \ran\ cross
sections from the Atomki-V2 AOMP are closer to the new experimental data
because these \ran\ cross sections are slightly higher than those from the other
AOMPs under study. The reason for the relatively low \ran\ cross sections from
the AVR AOMP is most likely the adjustment of the parameters -- among others
-- to the relatively low cross sections measured by Palumbo {\it et al.}\
\cite{Palumbo_PRC2012_a-n} for the \tetwonull \ran \xetwoiii\ reaction. There
is no obvious explanation for the more or less pronounced underestimation of
the \ran\ cross sections by the DEM1, DEM2, and DEM3 AOMPs. An undocumented
modification of DEM1, DEM2, and DEM3 AOMPs in TALYS-1.96 sharpens the
deviations to the new experimental data.

In general, the calculation of \ran\ cross sections is mainly sensitive to the
chosen AOMP (as explained in Sec.\,\ref{sec:gen}, \ref{sec:stat},
\ref{sec:technical}, and in
Fig.\,\ref{fig:te120a_branching}). Other ingredients of the statistical model
calculations play only a relatively minor role. Thus, further investigations
beyond the AOMP were only necessary in two cases which are the \tetwonull \ran
\xetwoiii\ and the \tethreenull \ran \xethreeiii\ reactions.

For \tetwonull \ran \xetwoiii , there is some tension between our new
experimental data and the data by Palumbo {\it et al.}
\cite{Palumbo_PRC2012_a-n} close above the threshold. In this energy region,
the contributions of the \ran\ and \rag\ channels have the same order of
magnitude. Whereas the new experimental data are nicely reproduced by the
Atomki-V2 AOMP and TALYS default settings otherwise, the Palumbo data require
an enhanced \rag\ and reduced \ran\ channel which can be achieved e.g.\ by an
enhancement of the GSF. Further experimental data close to the \ran\ threshold
are required to resolve this issue (see Fig.\,\ref{fig:te120an_test}).

For \tethreenull \ran \xethreeiii\ we find that the combination of the
Atomki-V2 AOMP with default parameters underestimates the new \ran\
data at higher energies. However, as the Atomki-V2 AOMP reproduces the total
cross section \stot\ (as determined by elastic scattering), the reason for the
underestimation of the \ran\ cross sections must be related to other
ingredients of the statistical model. It is found that the \ran\ data can be
nicely described by an enhancement of the LD in the residual \xethreeiii\
nucleus which favors the \ran\ channel and reduces the otherwise far
dominating \rann\ channel at energies above 14 MeV (see
Fig.\,\ref{fig:te130an_test}).

In general, it has to be pointed out that additional experimental data for
other reaction channels like \rag\ or \rap\ or \rann\ or isomer branchings are
required to constrain the other ingredients of the statistical model beyond
the AOMP. As the availability of such additional data is very limited, the
other ingredients of the statistical model like the GSF and the LD cannot be
well constrained.

Finally, we study the low-energy behavior of the AOMPs for the example of the
\tetwonull \rag \xetwoiv\ reaction. As already shown in
Fig.\,\ref{fig:te120a_branching}, at low energies below the \ran\ threshold
around 10 MeV, the \rag\ cross section is practically identical to the total
cross section \stot . Thus, for simplicity we investigate only \stot\ in the
following. The result is shown in Fig.\,\ref{fig:te120a_total_low} down to deep
sub-barrier energies.
\begin{figure}
\centering
\includegraphics[width=\columnwidth]{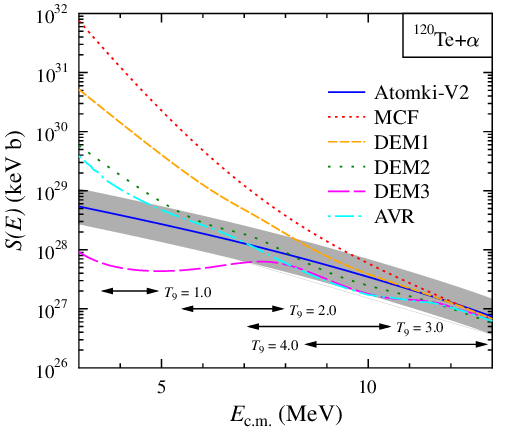}
\caption{
  \label{fig:te120a_total_low}
  Total cross section \stot\ of \tetwonull\ + \al\ at low energies (shown as
  astrophysical \sfact s). The claimed uncertainty of the Atomki-V2 AOMP is
  indicated by the grey-shaded area. The horizontal arrows indicate the
  classical Gamow windows for $T_9 = 1$, 2, 3, and 4. Further discussion see
  text.
  }
\end{figure}

Whereas at higher energies the Atomki-V2 AOMP provides the highest \stot , the
other AOMPs show much higher \stot\ towards very low energies. Below about 10
MeV most of the other AOMPs (except the DEM3 AOMP) show a more or less
pronounced steep increase of the \sfact\ towards lower energies. This steep
increase is typically related to the tail of the imaginary part of the AOMP at
large radii beyond the colliding nuclei which is not well-constrained (for a
detailed discussion see \cite{Szucs_NPA9_au197} and the Supplement of
\cite{Mohr_PRL2020}). At 10 MeV the predicted total cross sections vary by
about a factor of 2.4 between the AVR, DEM1, DEM2, DEM3, and Atomki-V2
AOMPs. (We exclude the MCF AOMP from the discussion because the dramatic
overestimation of 
low-energy \sfact s is well-known and confirmed here.) The range of
predictions from the AVR, DEM1, DEM2, DEM3, and Atomki-V2 AOMPs increases to a
factor of 9.5 at 7.5 MeV and 84.9 at 5.0 MeV. Thus, huge discrepancies between
the predictions from various AOMPs re-appear also in the present study of \al
-induced reactions on tellurium isotopes, but only at energies far below the
Coulomb barrier.

Fig.\,\ref{fig:te120a_total_low} also shows the claimed uncertainty of a factor
of two for the Atomki-V2 AOMP. It is interesting to note that the AVR and DEM2
AOMPs remain within this uncertainty range at all energies above about 5
MeV. For the DEM1 and DEM3 AOMPs this holds above 8 MeV and 7 MeV,
respectively.

Because of the overall good description of the present \ran\ cross sections
from the Atomki-V2 AOMP, we do not provide new astrophysical reaction rates in
this paper. Instead, we recommend to use the reaction rates from the Atomki-V2
AOMP for all tellurium isotopes which are published in \cite{Mohr_ADNDT2021}
and available for numerical download.

%\clearpage

\section{Summary and conclusions}
\label{sec:conc}
In the present work, the \ran\ cross section of four Te isotopes ($^{120,122,124,130}$Te) was measured between 10 and 17\,MeV $\alpha$ energies using the activation technique. For these reactions experimental data were not at all available in the literature so far, or not in the presently studied energy range and with sufficiently high precision. In the case of the $^{130}$Te\ran$^{133}$Xe reaction the cross section leading to the ground and isomeric states of $^{133}$Xe were measured separately. 

Since the cross sections were measured close above the \ran\ thresholds, the results allow the study of the $\alpha$-nucleus optical potential, which is an important quantity in various nuclear astrophysics models. The new experimental data were compared with statistical model predictions using various AOMP models. It is found that the recently developed Atomki-V2 potential gives the best description of the experimental data. This result provides a further proof of the generally good predictive power of this potential in this mass and energy range. The application of the recently compiled reaction rates based on this potential \cite{Mohr_ADNDT2021} is thus recommended in nucleosynthesis calculations. 

\begin{acknowledgments}
This work was supported by NKFIH grants No. NN128072, K134197 and K138031 and by the \'UNKP-23-3-I-DE-165 New National Excellence Programs of the Ministry for Culture and Innovation of Hungary. The financial support of the Hungarian Academy of Sciences (Infrastructure grants), and the Economic Development and Innovation Operational Programme (GINOP-2.3.3–15-2016-00005) grant, co-funded by the EU, is also acknowledged.
\end{acknowledgments}

\end{document}